\documentclass[12pt]{article}
\usepackage{graphicx}
\usepackage[cp1251]{inputenc}
\usepackage[russian]{babel}
\usepackage{rotating}
\begin{document}
 \noindent {\footnotesize\it Astronomy Letters, 2017, Vol. 43, No 4, pp. 241--251}
 \newcommand{\dif}{\textrm{d}}

 \noindent
 \begin{tabular}{llllllllllllllllllllllllllllllllllllllllllllll}
 & & & & & & & & & & & & & & & & & & & & & & & & & & & & & & & & & & & & & \\\hline\hline
 \end{tabular}

  \vskip 0.5cm
 \centerline{\bf\large Refinement of the Parameters of Three Selected Model Galactic}
 \centerline{\bf\large Potentials Based on the Velocities of Objects }
 \centerline{\bf\large at Distances up to 200 kpc}
 \bigskip
 \bigskip
  \centerline
 {
 V.V. Bobylev$^1$,
 A.T. Bajkova$^1$, and
 A.O. Gromov$^2$
 }
 \bigskip
{\small \it

 (1) Pulkovo Astronomical Observatory, Russian Academy of Sciences,
     Pulkovskoe sh. 65, St. Petersburg, 196140 Russia

 (2) Department of Space Technologies and Applied Astrodynamics,
 St. Petersburg State University, Universitetskii pr. 35, St. Petersburg, 198504 Russia
 }
 \bigskip
 \bigskip
 \bigskip

 {
{\bf Abstract}---This paper is a continuation of our recent paper
devoted to refining the parameters of three component (bulge,
disk, halo) axisymmetric model Galactic gravitational potentials
differing by the expression for the dark matter halo using the
velocities of distant objects. In all models the bulge and disk
potentials are described by the Miyamoto--Nagai expressions. In
our previous paper we used the Allen--Santill\'an (I),
Wilkinson--Evans (II), and Navarro--Frenk--White (III) models to
describe the halo. In this paper we use a spherical logarithmic
Binney potential (model IV), a Plummer sphere (model V), and a
Hernquist potential (model VI) to describe the halo. A set of
present-day observational data in the range of Galactocentric
distances $R$ from 0 to 200 kpc is used to refine the parameters
of the listed models, which are employed most commonly at present.
The model rotation curves are fitted to the observed velocities by
taking into account the constraints on the local matter density
$\rho_\odot=0.1M_\odot$ pc$^{-3}$ and the force $K_{z=1.1}/2\pi
G=77M_\odot$ pc$^{-2}$ acting perpendicularly to the Galactic
plane. The Galactic mass within spheres of radius 50 and 200 kpc
are shown to be, respectively,
  $M_{50}=(0.409\pm0.020)\times10^{12}M_\odot$ and
 $M_{200}=(1.395\pm0.082)\times10^{12}M_\odot$ in model IV,
  $M_{50}=(0.417\pm0.034)\times10^{12}M_\odot$ and
 $M_{200}=(0.469\pm0.038)\times10^{12}M_\odot$ in model V, and
  $M_{50}=(0.417\pm0.032)\times10^{12}M_\odot$ and
 $M_{200}=(0.641\pm0.049)\times10^{12}M_\odot$ in model VI. Model VI
looks best among the three models considered here from the
viewpoint of the achieved accuracy of fitting the model rotation
curves to the measurements. This model is close to the
Navarro--Frenk--White model III refined and considered best in our
previous paper, which is shown using the integration of the orbits
of two globular clusters, Lynga~7 and NGC~5053, as an example.
  }

\section*{INTRODUCTION}
It is necessary to have a reliable model gravitational potential
to study the structure of the Galaxy and its dynamical properties
or to construct the orbits of Galactic objects. The circular
velocities of objects $V_{circ}$ at various distances $R$ from the
Galactic rotation axis (Galactic rotation curve) are among the
main sources for the construction of such a model. The accuracy of
any model potential depends on the accuracy of determining the
distances to the objects used for its construction.

At present, a high accuracy of distance measurements, on average,
$\sim$10\%, has been achieved for a number of Galactic objects. In
particular, these include more than 100 Galactic masers with
measured trigonometric parallaxes, line-of-sight velocities, and
proper motions (Reid et al. 2014). Hydrogen clouds at the tangent
points also have a high positional accuracy. However, all these
objects are no farther than $R=20$~kpc. The Galactic rotation
curve constructed from hydrogen clouds and masers has been used by
various authors to refine several model potentials (Irrgang et al.
2013; Bobylev and Bajkova 2013; Reid et al. 2014; Butenko and
Khoperskov 2014). This rotation curve is characterized by a fairly
large circular velocity of the Galaxy at the solar distance,
$V_\odot\sim250$~km s$^{-1}$. This leads to an increase in the
estimated mass of the Galaxy and its components (disk, halo)
compared to that obtained at the standard velocity
$V_\odot=220$~km s$^{-1}$.

Analysis of the observational data at progressively larger $R$ is
of great interest. The possibilities for this are continuously
expanding. Here is an incomplete list of the most interesting
results. Gnedin et al. (2010) found the Galactic mass within a
sphere of radius 80 kpc from stars at distances $R<80$~kpc to be
$M_{80}=0.69^{+3.0}_{-1.2}\times10^{12}M_\odot$. Deason et al.
(2012b) found the Galactic mass within a sphere of radius 150~kpc
from distant blue horizontal branch giants to be
$M_{150}=(0.5-0.1)\times10^{12}M_\odot$. Bhattacharjee et al.
(2014) estimated the Galactic mass within a sphere of radius
200~kpc from combined data to be
$M_{200}\geq(0.68\pm0.41)\times10^{12}M_\odot$. Eadie et al.
(2015) found the Galactic mass within a sphere of radius 260~kpc
from globular clusters and dwarf galaxies to be
$M_{260}=1.37\times10^{12}M_\odot$. Based on the effect of local
Hubble flow deceleration and data on galaxies in the neighborhoods
of the Local Group, Karachentsev et al. (2009) estimated the total
mass of the Local Group and the ratio of the Galactic and M31 mass
to be 4:5. The total mass of the Galaxy obtained by this
independent method is $M_{350}=(0.84\pm0.09)\times10^{12}M_\odot$.

This article is a continuation of the paper by Bajkova and Bobylev
(2016) devoted to refining the parameters of model Galactic
gravitational potentials, where three-component (bulge, disk,
halo) models differing by the expression for the dark matter halo
were considered. The expressions from Miyamoto and Nagai (1975)
were traditionally used to describe the bulge and the disk, and
the following three models were considered to describe the halo:
the Allen--Santill\'an (I), Wilkinson--Evans (II), and
Navarro--Frenk--White (III) models.

The goal of this paper is to refine the parameters of the Galactic
gravitational potential for other models of the dark matter halo.
For this purpose, we use three models where the halo is
represented by a logarithmic Binney potential (Binney 1981; Evans
1993; Fellhauer et al. 2006) (model IV), a Plummer (1911) sphere
(model V), and a Hernquist (1990) potential (model VI). Here, as
in our previous paper (Bajkova and Bobylev 2016), we use
observational data in the range of distances 0--200~kpc and
constraints on the local matter density $\rho_\odot$ and the force
$K_{z=1.1}$. The selected models are employed most commonly at
present, because they are realistic and mathematically simple and
their analytic expressions are closed. Therefore, their correction
based on new data is an important task.

 \section*{DATA}\label{data}
Since this is a continuation of our previous paper (Bajkova and
Bobylev 2016), we use exactly the same data. First, these include
the line-of-sight velocities of HI clouds at the tangent points
from Burton and Gordon (1978). These data on the rotation curve
fill the range of distances $R<4$~kpc. Second, these include a
sample of masers with measured trigonometric parallaxes, proper
motions, and line-of-sight velocities. They are located in the
range of distances $R$ from 4 to 20 kpc. Third, these include the
average circular rotation velocities from Bhattacharjee et al.
(2014) calculated using objects at distances $R$ from 20 to
$\approx$200 kpc. The velocities of 1457 blue horizontal branch
giants, 2227 K giants, 16 globular clusters, 28 distant halo
giants, and 21 dwarf galaxies were used for this purpose. Since
Bhattacharjee et al. (2014) constructed the Galactic rotation
curve with $R_\odot=8.3$~kpc and $V_\odot=244$ km s$^{-1}$, we
also calculate the circular velocities of objects with these
parameters.

The following can be said in more detail about the masers. The
VLBI measurements of 103 masers are described in the general
review of Reid et al. (2014). Subsequently, these authors
published improved values of the measured parameters for some of
the masers (Wu et al. 2014; Choi et al. 2014; Sato et al. 2014;
Hachisuka et al. 2015). After the addition of the most recent
published VLBI measurements (Motogi et al. 2015; Burns et al.
2015), we obtained a sample with data on 130 sources. We do not
include the masers located at distances $R<4$~kpc in our sample,
because their velocities are distorted by the central Galactic
bar.

As a result, we rely on the line-of-sight velocities of hydrogen
clouds at the tangent points and the data on 130 masers with
measured trigonometric parallaxes at distances $R$ less than
25~kpc, and the rotation curve from Bhattacharjee et al. (2014)
serves us as the data at greater distances.

\section*{MODEL GALACTIC POTENTIALS}
\subsection*{Introductory Mathematical Expressions}
In all of the models being considered here the axisymmetric
Galactic potential is represented as a sum of three components~---
a central spherical bulge $\Phi_b(r(R,z))$, a disk
$\Phi_d(r(R,z))$, and a massive spherical dark matter halo
$\Phi_h(r(R,z))$:
 \begin{equation}
 \renewcommand{\arraystretch}{2.8}
  \Phi(R,z)=\Phi_b(r(R,z))+\Phi_d(r(R,z))+\Phi_h(r(R,z)).
 \label{pot}
 \end{equation}
Here, we use a cylindrical coordinate system ($R,\psi,z$) with the
coordinate origin at the Galactic center. In a rectangular
coordinate system $(x,y,z)$ with the coordinate origin at the
Galactic center the distance to a star (spherical radius) will be
$r^2=x^2+y^2+z^2=R^2+z^2$.

In accordance with the convention adopted in Allen and Santill\'an
(1991), we express the gravitational potential in units of 100
km$^2$ s$^{-2},$ the distances in kpc, and the masses in units of
the Galactic mass $M_{gal}=2.325\times10^7 M_\odot$, corresponding
to the gravitational constant $G=1.$

The expression for the mass density follows from the Poisson
equation
\begin{equation}
  \renewcommand{\arraystretch}{1.6}
4\pi G\rho(R,z)=\nabla^2\Phi(R,z) \label{pois1}
\end{equation}
and is
 \begin{equation}
 \begin{array}{lll}
 \renewcommand{\arraystretch}{2.8}
 \displaystyle
 \rho(R,z)=\frac{1}{4\pi G}\Bigg{(}\frac{d^2\Phi(R,z)}{dR^2}+
           \frac{1}{R}\frac{d\Phi(R,z)}{dR}+\frac{d^2\Phi(R,z)}{dz^2}\Bigg{)}.
    \label{pois2}
 \end{array}
\end{equation}
The force acting in the $z$ direction perpendicular to the
Galactic plane is defined as
\begin{equation}
 \renewcommand{\arraystretch}{1.2}
  K_z(z,R)=-\frac{d\Phi(z,R)}{dz}.
 \label{Kz}
\end{equation}
We will need Eqs. (3) and (4) below to solve the problem of
fitting the parameters of the model gravitational potentials with
the constraints imposed on the local dynamical matter density
$\rho_\odot$ and the force $K_z(z,R_\odot)$ at $z=1.1$~kpc that
are known from observations.

In addition, we will need the expressions to calculate:

(1)~the circular velocities
\begin{equation}
 V_{circ}(R)=\sqrt{R\frac{d\Phi(R,0)}{dR}},
 \label{V}
\end{equation}

(2)~the Galactic mass contained in a sphere of radius $r$ based on
the spherically symmetric model
\begin{equation}
 M(r)=r^2\frac{d\Phi(r)}{dr},
 \label{m}
\end{equation}

(3)~the parabolic velocity or the escape velocity of a star from
the attractive Galactic field
\begin{equation}
 V_{esc}(R,z)=\sqrt{-2\Phi(R,z)},
 \label{Vesc}
\end{equation}

(4)~the Oort parameters
\begin{equation}
 A=\frac{1}{2}\Omega_\odot^{'}R_\odot,
 \label{A}
\end{equation}
\begin{equation}
B=\Omega_\odot+A, \label{B}
\end{equation}
where $\Omega=V/R$ is the angular velocity of Galactic rotation
$(\Omega_\odot=V_\odot/R_\odot)$, $\Omega^{'}_\odot$ is the first
derivative of the angular velocity with respect to $R$ and
$R_\odot$ is the Galactocentric distance of the Sun;

(5)~the surface density of gravitating matter within $z_{out}$ of
the Galactic $z=0$ plane
 \begin{equation}
 \begin{array}{lll}
 \renewcommand{\arraystretch}{2.8}
 \displaystyle
 \Sigma_{out}(z_{out})=  2\int_0^{z_{out}} \rho(R,z)dz=
       \frac{K_{z}}{2\pi G}+\frac{2z_{out}(B^2-A^2)}{2\pi G},
 \label{Sigma}
 \end{array}
\end{equation}
Note that the spherically symmetric model describes satisfactorily
the density distribution in a galaxy only when the disk mass is
much smaller than the mass of its spherical components. This
condition is fulfilled in the case of our Galaxy. However, even if
this condition is not fulfilled completely, Eq. (6) is well suited
for a rough estimation of the total mass within a radius $r.$
Generally, the mass found within the spherically symmetric model
will be an upper limit: $M(r)\le r^2 d\Phi(r)/dr$ (Zasov and
Postnov 2006).

\subsection*{Bulge and Disk}
In all of the models being considered here the bulge, $\Phi_b(r)$
and disk, $\Phi_d(r(R,z))$ potentials are represented by the
expressions from Miyamoto and Nagai (1975):
 \begin{equation}
 \renewcommand{\arraystretch}{1.2}
  \Phi_b(r)=-\frac{M_b}{(r^2+b_b^2)^{1/2}},
  \label{bulge}
 \end{equation}
 \begin{equation}
 \Phi_d(R,z)=-\frac{M_d}{\{R^2+[a_d+(z^2+b_d^2)^{1/2}]^2\}^{1/2}},
 \label{disk}
\end{equation}
where $M_b$ and $M_d$ are the masses of the components, $b_b,$
$a_d,$ and $b_d$ are the scale lengths of the components in kpc.
The corresponding expressions for the mass densities $\rho_b(R,z)$
and $\rho_d(R,z)$ are
\begin{equation}
\rho_b(r)=\frac{3b_b^2 M_b}{4\pi G(r^2+b_b^2)^{5/2}}, \label{ro-b}
\end{equation}
 \begin{equation}
 \begin{array}{lll}
 \renewcommand{\arraystretch}{2.8}
 \displaystyle
 \rho_d(R,z)=\frac{b_d^2 M_d}{4\pi G(z^2+b_d^2)^{3/2}}
             \frac{a_d R^2+(a_d+3\sqrt{z^2+b_d^2})(a_d+\sqrt{z^2+b_d^2})^2}
 {(R^2+(a_d+\sqrt{z^2+b_d^2})^2)^{5/2}}.
  \label{ro-d}
 \end{array}
\end{equation}
Expressions (11) and (13) describe a Plummer (1911) sphere, while
relations (12) and (14) describe a generalized Kuzmin (1956) disk.

Integrating the mass densities over the entire volume of the
Galaxy gives the expected bulge and disk masses: $m_b=M_b,$ and
$m_d=M_d$. The bulge and disk contributions to the circular
velocity are, respectively,
\begin{equation}
 \renewcommand{\arraystretch}{1.2}
 V_{circ(b)}^2(R)=\frac{M_b R^2}{(R^2+b_b^2)^{3/2}}, \label{Vc-b}
 \end{equation}
\begin{equation}
 V_{circ(d)}^2(R)=\frac{M_d R^2}{(R^2+(a_d+b_d)^2)^{3/2}}.
\label{Vc-d}
\end{equation}
The corresponding expressions for $K_z^b(z,R)$ and $K_z^d(z,R)$
are
\begin{equation}
  K_{z}^{b}(z,R)=\frac{z M_b}{(R^2+z^2+b_b^2)^{3/2}},
\label{Kz-b}
\end{equation}
 \begin{equation}
 \begin{array}{lll}
 \renewcommand{\arraystretch}{2.6}
 \displaystyle
  K_{z}^{d}(z,R)=\frac{z M_d (a_d+\sqrt{z^2+b_d^2})}
 {\sqrt{z^2+b_d^2}(R^2+(a_d+\sqrt{z^2+b_d^2})^2)^{3/2}}.
 \label{Kz-d}
 \end{array}
\end{equation}

\subsection*{Dark Matter Halo}
{\bf Model IV.} The halo component is represented by a logarithmic
potential in the form proposed by Binney (1981):
 \begin{equation}
 \renewcommand{\arraystretch}{1.2}
  \Phi_h(R,z)=-\frac{v^2_0}{2}\ln\Biggl(R^2+a^2_h+\frac{z^2}{q^2_{\Phi}}\Biggr),
 \label{halo-IV}
 \end{equation}
where $v_0$ is the velocity in km s$^{-1}$, $q_\Phi$ is the axial
ratio of the ellipsoid: $q_\Phi=1$ for a spherical halo,
$q_\Phi<1$ for an oblate one, and $q_\Phi>1$ for a prolate one. We
take $q_\Phi=1.$ In this case, the mass density is
\begin{equation}
  \rho_h(r)= \frac{v^2_0}{4\pi G} \frac{(3a^2_h+r^2)} {(r^2+a^2_h)^2}.
 \label{ro-h-IV}
 \end{equation}
It is easy to show that integrating the mass density over a sphere
of radius $r$ gives a halo mass that depends almost linearly on
$r.$ The contribution to the circular velocity is
 \begin{equation}
  V^2_{circ(h)}= \frac{v_0^2 R^2}{R^2+a^2_h}.
 \label{Vc-h-IV}
 \end{equation}
The expression for $K_z^h(z,R)$ is
\begin{equation}
K_z^h(z,R)=\frac{z v_0^2}{R^2+z^2+a_h^2}. \label{Kz-h-IV}
\end{equation}
In the three-component model Galactic potential of Fellhauer et
al. (2006) the Hernquist (1990) potential was used to describe the
bulge, the disk was described by Eq. (12), and the halo was
described by the logarithmic Binney (1981) potential (19). In the
original form this approach has been actively used in recent years
to construct the Galactic orbits of various objects (see, e.g.,
R{u}\u{z}i\u{c}ka et al. 2010; Gontcharov et al. 2011; Veras and
Evans 2013; Dremova et al. 2015).

{\bf Model V.} In this model we use a Plummer (1911) sphere
(coincident with Eq. (11)) to describe the halo potential. As a
result, we have
 \begin{equation}
  \Phi_h(r)=-\frac{M_h}{(r^2+a_h^2)^{1/2}}.
  \label{halo-V}
 \end{equation}
The mass density is
\begin{equation}
\rho_h(r)=\frac{3a_h^2 M_h}{4\pi G(r^2+a_h^2)^{5/2}}.
\label{ro-h-V}
\end{equation}
Integrating the mass density over the entire volume of the Galaxy
gives the mass $m_h=M_h$. The contribution to the circular
velocity is
\begin{equation}
V_{circ(h)}^2(R)=\frac{M_h R^2}{(R^2+a_h^2)^{3/2}}. \label{Vc-h-V}
\end{equation}
The expression for $K_z^h(z,R)$ is
\begin{equation}
K_{z}^{h}(z,R)=\frac{z M_h}{(R^2+z^2+a_h^2)^{3/2}}. \label{Kz-h-V}
\end{equation}
Such a three-component model Galactic potential (a bulge in the
form (11), a disk in the form (12), and a halo in the form (23))
was first proposed by Dauphole and Colin (1995). Therefore, we
call this approach the Dauphole--Colin model, and this approach is
much in demand at present. For example, it was used by Kalirai et
al. (2007) to study the orbit of the globular cluster NGC 6397, by
Just et al. (2009) to construct the Galactic orbits of open
clusters, and by Bailer-Jones (2015) to study close stellar
encounters.

{\bf Model VI.} The halo component is represented by the Hernquist
(1990) potential
 \begin{equation}
  \Phi_h(r)=-\frac{M_h}{r+a_h}.
  \label{halo-VI}
 \end{equation}
The mass density is
\begin{equation}
 \rho_h(r)=\frac{a_h M_h}{2\pi G r(r+a_h)^3}.
 \label{ro-h-VI}
\end{equation}
Integrating the mass density over the entire volume of the Galaxy
gives the mass $m_h=M_h$. The contribution to the circular
velocity is
\begin{equation}
 V_{circ(h)}^2(R)=\frac{M_h R}{(R+a_h)^2}.
 \label{Vc-h-VI}
\end{equation}
The expression for $K_z^h(z,r)$ is
\begin{equation}
 K_{z}^{h}(z,r)=\frac{z M_h}{r(r+a_h)^2}.
 \label{Kz-h-VI}
\end{equation}
The potential (27) is commonly and widely used to describe the
central spherical bulge (Fellhauer et al. 2006). However, this
potential is occasionally used to describe the halo or the
bulge--halo pair (see, e.g., Eadie et al. 2015; Capuzzo--Dolcetta
and Fragione 2015).

\subsection*{Parameter Fitting}
The parameters of the model potentials are found by least-squares
fitting to the measured rotation velocities of Galactic objects.
We use the unit weights, because they provide the smallest
absolute residual between the data and the model rotation curve.
In addition, we used (Irrgang et al. 2013) the constraints on (1)
the local dynamical matter density
$\rho_\odot=0.1M_\odot$~pc$^{-3}$ and (2) the force acting
perpendicularly to the Galactic plane or, more specifically,
$K_{z=1.1}/2р G=77M_\odot$ pc$^{-2}$.

The local dynamical matter density $\rho_\odot$, which is the sum
of the bulge, disk, and dark matter densities in a small solar
neighborhood, together with the surface density $\Sigma_{1.1}$ are
the most important additional constraints in the problem of
fitting the parameters of the model potentials to the measured
circular velocities (Irrgang et al. 2013):
\begin{equation}
\rho_\odot=\rho_b(R_\odot)+\rho_d(R_\odot)+\rho_h(R_\odot),
\label{ro}
\end{equation}
\begin{equation}
\Sigma_{1.1}=
  \int\limits^{1.1\,\hbox {\footnotesize\it kpc}}_{-1.1\,\hbox {\footnotesize\it kpc}}
(\rho_b(R_\odot,z)+\rho_d(R_\odot,z)+\rho_h(R_\odot,z))dz.
\label{Sig}
\end{equation}
The surface density is closely related to the force $K_z(z,R)$ in
accordance with Eq. (10). Since the two most important parameters
$\rho_\odot$ and $K_z/2\pi G$ are known from observations with a
sufficiently high accuracy, introducing additional constraints on
these two parameters allows the parameters of the gravitational
potential to be refined significantly.

As a result, the parameter fitting problem was reduced to
minimizing the following quadratic functional $F:$
 \begin{equation}
 \begin{array}{lll}
 \renewcommand{\arraystretch}{2.6}
 \displaystyle
 \min F=\sum_{i=1}^N
 (V_{circ}(R_i)-\widetilde{V}_{circ}(R_i))^2+
        \alpha_1(\rho_\odot-\widetilde{\rho}_\odot)^2+\alpha_2(K_{z=1.1}/2\pi G-
 \widetilde{K}_{z=1.1}/2\pi G)^2,
  \label{F}
 \end{array}
\end{equation}
where the measured quantities are denoted by the tilde, $R_i$ are
the distances of the objects with measured circular velocities,
$\alpha_1$ and $\alpha_2$ are the weight factors at the additional
constraints that were chosen so as to minimize the residual
between the data and the model rotation curve provided that the
additional constraints hold with an accuracy of at least 5\%.
Based on the constructed models, we calculated the local surface
density of the entire matter $\rho_\odot$ and $K_{z=1.1}/2\pi G$
related to $\Sigma_{1.1}$ and $\Sigma_{out}.$

 {\begin{table}[t]                            
 \caption[] {\small\baselineskip=1.0ex
 The model parameters found by fitting to the data}
 \label{t:1}
 \small \begin{center}\begin{tabular}{|c|c|c|c|c|}\hline
        Parameters &    model~IV       &    model~V        & model VI          \\\hline
  $M_b$ ($M_{g}$) &    $486\pm10$      &    $456\pm40$     & $ 461\pm22$        \\
  $M_d$ ($M_{g}$) &    $3079\pm23$     &   $3468\pm71$     & $ 2950\pm33$       \\
  $M_h$ ($M_{g}$) & $14210\pm858~(^*)$ &  $16438\pm1886$   & $29677\pm2791$     \\
  $m_b$ ($10^9 M_\odot$) &    $11.30\pm0.23$      &    $10.60\pm0.93$     & $ 10.71\pm0.51$        \\
  $m_d$ ($10^{10} M_\odot$) &    $7.16\pm0.05$     &   $8.06\pm0.16$     & $ 6.86\pm0.07$       \\
  $m_h$ ($10^{11} M_\odot$) & $13.12\pm0.20~(^{**})$ &  $3.81\pm0.44$   & $6.88\pm0.65$     \\
      $b_b$ (kpc) &  $0.2769\pm0.0052$ & $0.2647\pm0.0057$ & $ 0.2720\pm0.0130$ \\
      $a_d$ (kpc) &    $3.54\pm0.06$   &   $2.94\pm0.076$  & $ 3.85\pm0.08$     \\
      $b_d$ (kpc) &  $0.2997\pm0.0023$ & $0.3126\pm0.0022$ & $ 0.3090\pm0.0010$ \\
      $a_h$ (kpc) &    $3.20\pm0.45$   &  $16.57\pm1.38$   & $21.27\pm1.06$     \\\hline
 Residual noise entropy $E$    &  -29.11   &  -29.72  &  -24.96\\\hline $\delta $ (km s$^{-1}$)         &     15.04 &   14.89  &   13.23
 \\\hline
 \end{tabular}\end{center}
 {\small The Galactic mass unit is $M_{g}=2.325\times 10^7 M_\odot$,
   $(^*)$ $v_0^2/2$ in km$^2$s$^{-2}$ is given here,
   $(^{**})$ the halo mass within a radius of 200 kpc is given here.}
 \end{table}}
 {\begin{table}[t]                            
 \caption[] {\small\baselineskip=1.0ex
 The quantities calculated from the parameters of the model potentials found}
 \label{t:2}
 \small \begin{center}\begin{tabular}{|l|r|r|r|r|r|r|}\hline
                              Parameters &       model~IV &    model~V &    model~VI \\\hline
  $(\rho_\odot)_d$ ($M_\odot$ pc$^{-3}$) & $0.092\pm0.009$ & $0.089\pm0.010$ & $0.090\pm0.010$\\
  $(\rho_\odot)_h$ ($M_\odot$ pc$^{-3}$) & $0.008\pm0.001$ & $0.011\pm0.001$ & $0.011\pm0.001$\\
  ~~~ $\rho_\odot$ ($M_\odot$ pc$^{-3}$) & $0.100\pm0.010$ & $0.100\pm0.010$ & $0.100\pm0.010$\\
 $K_{z=1.1}/2\pi G$  ($M_\odot$ pc$^{-2}$) &    $77.0\pm6.3$ & $77.1\pm6.6$    & $77.2\pm5.8$\\
  ~   $\Sigma_{1.1}$ ($M_\odot$ pc$^{-2}$) &    $71.4\pm6.4$ & $78.6\pm7.9$    & $76.9\pm6.4$\\
  ~   $\Sigma_{out}$ ($M_\odot$ pc$^{-2}$) &    $45.2\pm7.1$ & $75.0\pm14.2$  & $68.9\pm10.1$\\
  $V_{esc, ~R=R_\odot}$ (km s$^{-1}$)                           & $450.2\pm 8.6$ & $516.0\pm21.4$ & $524.8\pm18.2$\\
  $V_{esc, ~R=200\, kpc}$ (km s$^{-1}$) & $550.7\pm16.7$ & $142.5\pm5.7$ & $173.9\pm6.8$\\
   $V_\odot$ (km s$^{-1}$) & $ 241.3\pm3.9$  & $ 238.8\pm9.4$ & $243.1\pm6.8$\\
  $A$ (km s$^{-1}$ kpc$^{-1}$) & $ 16.10\pm0.62$ & $ 14.49\pm0.60$ & $15.05\pm0.52$\\
  $B$ (km s$^{-1}$ kpc$^{-1}$) & $-12.97\pm0.69$ & $-14.27\pm1.15$ & $-14.24\pm0.84$\\
 $M_{G_{(R\leq~50\, kpc)}}$ ($10^{12}M_\odot$) & $0.409\pm0.020$ & $0.417\pm0.034$ & $0.417\pm0.032$\\
 $M_{G_{(R\leq100\, kpc)}}$ ($10^{12}M_\odot$) & $0.738\pm0.040$ & $0.457\pm0.037$ & $0.547\pm0.042$\\
 $M_{G_{(R\leq150\, kpc)}}$ ($10^{12}M_\odot$) & $1.066\pm0.061$ & $0.466\pm0.037$ & $0.607\pm0.047$\\
 $M_{G_{(R\leq200\, kpc)}}$ ($10^{12}M_\odot$) & $1.395\pm0.082$ & $0.469\pm0.038$ & $0.641\pm0.049$\\
 \hline
 \end{tabular}\end{center}
 \end{table}}
\begin{figure}[p]
{\begin{center}
   \includegraphics[width=0.99\textwidth]{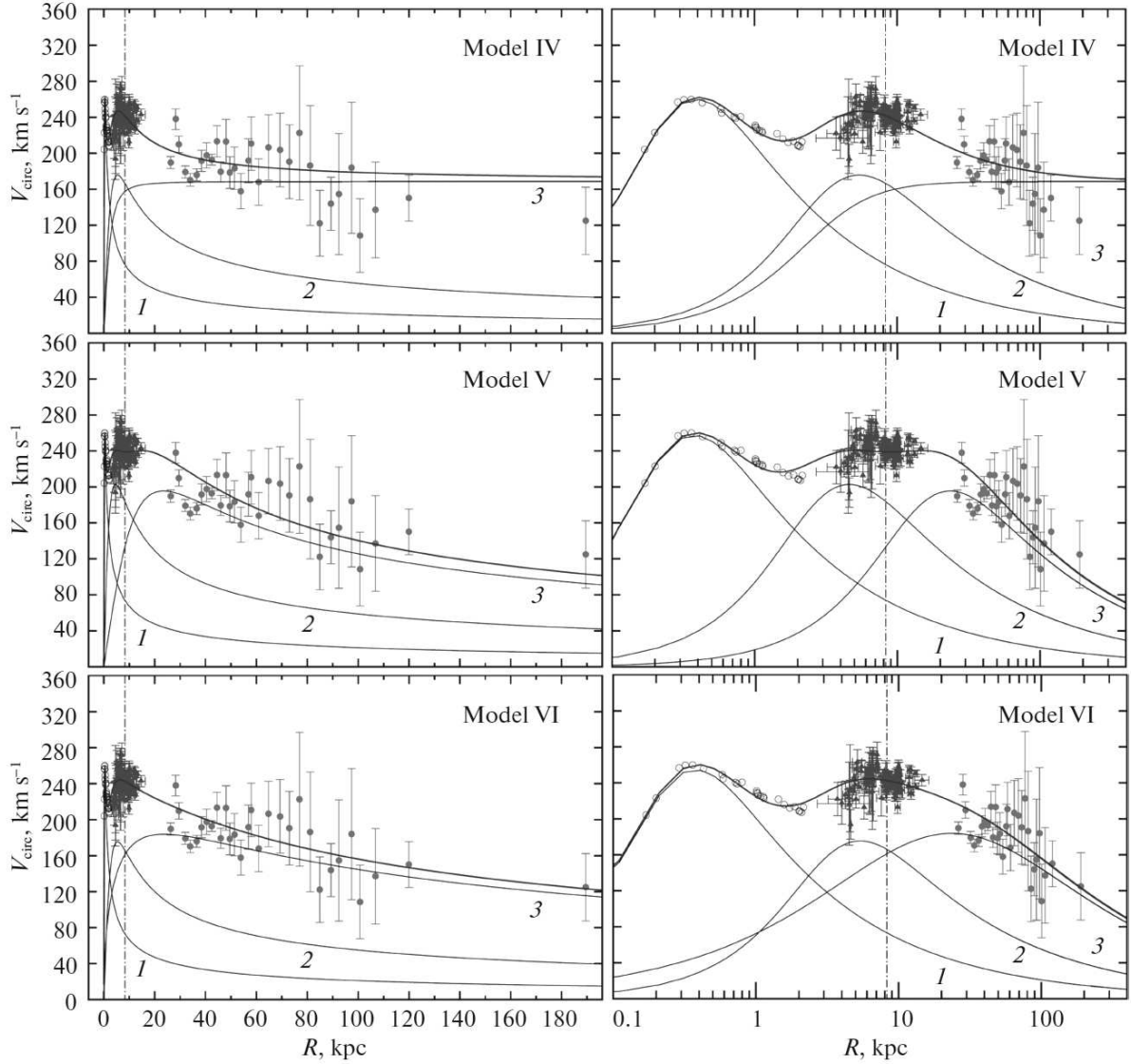}
 \caption{\small
Model Galactic rotation curves $V_{circ}(R)$ for potentials IV, V,
and VI in the linear (left) and logarithmic (right) distance
scales as well as the measured rotation velocities of objects
$\widetilde{V}_{circ}(R_i)$; the vertical dash–dotted line marks
the Sun's position, numbers 1, 2, and 3 denote the bulge, disk,
and halo contributions, respectively; the open circles, filled
triangles, and filled circles indicate the HI velocities, the
velocities of masers with measured trigonometric parallaxes, and
the velocities from Bhattacharjee et al. (2014), respectively.
  } \label{f2b}
\end{center}}
\end{figure}
\begin{figure}[t]
{\begin{center}
   \includegraphics[width=0.45\textwidth]{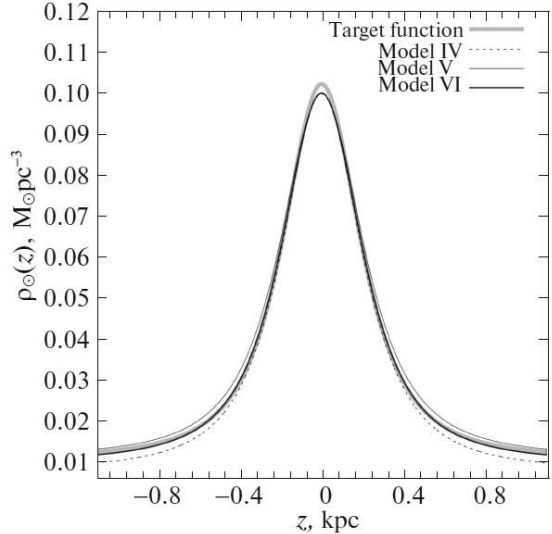}
 \caption{\small
Results of fitting $\rho_\odot(z)$ for models IV, V, and VI. The
functions from Irrgang et al. (2013) are considered as the target
functions $\rho_\odot(z)$.
  } \label{frho}
\end{center}}
\end{figure}
\begin{figure}[t]
{\begin{center}
   \includegraphics[width=0.8\textwidth]{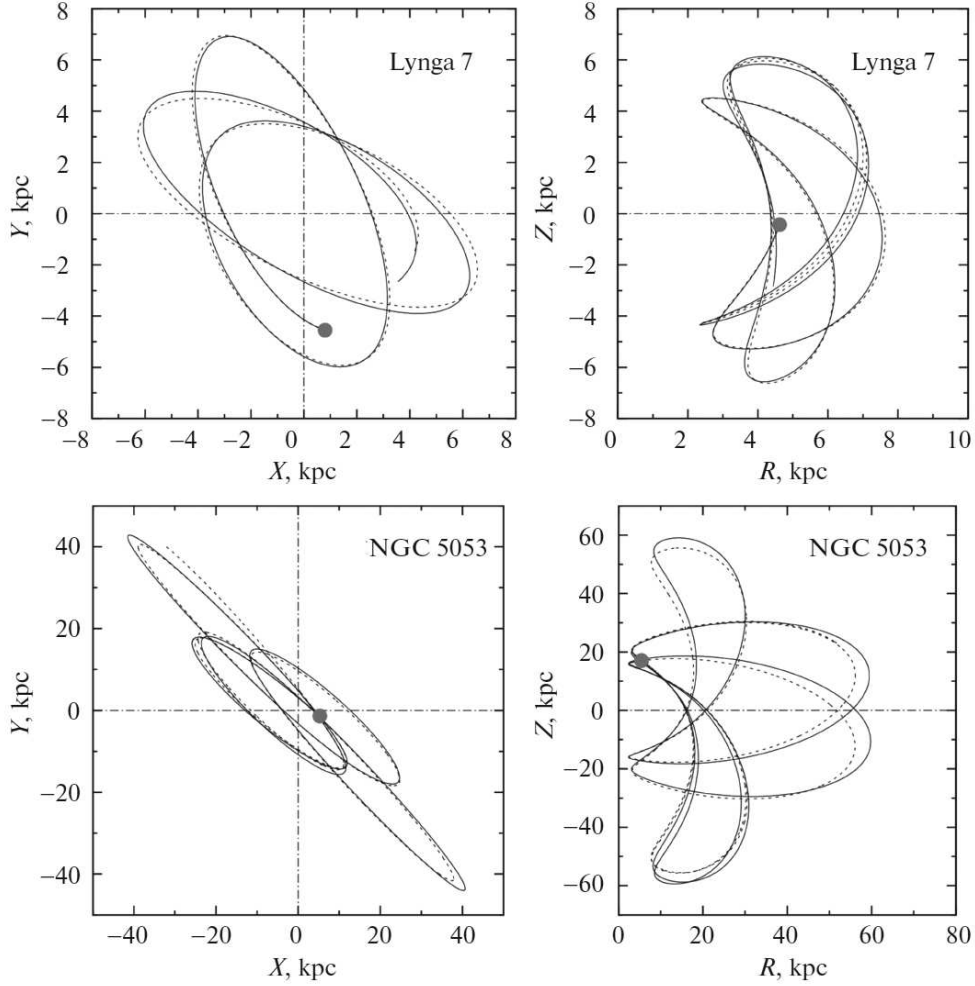}
 \caption{\small
The Galactic orbits of the globular clusters Lynga 7 (top) and NGC
5053 (bottom) constructed using models III (solid line) and VI
(dotted line); the gray circle marks the current position of the
cluster.
  } \label{fig3}
\end{center}}
\end{figure}
 {\begin{table}[t]                                    
 \caption[]
 {\small\baselineskip=1.0ex
 The characteristics of the Galactic orbits of the
 two selected globular clusters calculated using models III and VI
}
 \label{t:ecc}
 \small\baselineskip=1.0ex
\begin{center}\begin{tabular}{|l|c|c|c|c|c|c|c|}\hline
 Cluster & model & $\langle a_{min}\rangle,$ kpc & $\langle a_{max}\rangle,$ kpc & $\langle e\rangle$ \\\hline

 Lynga~7   &  III   & 4.34  & 8.00  & 0.30 \\
           &   VI   & 4.34  & 8.08  & 0.30 \\\hline
 NGC 5053  &  III   & 14.79 & 61.09 & 0.61 \\
           &   VI   & 14.80 & 57.93 & 0.59 \\\hline
 \end{tabular}\end{center}\end{table}}

 \section*{RESULTS AND DISCUSSION}
Table 1 provides the values of the seven sought for parameters
($M_b, M_d, M_h, b_b, a_d, a_b, a_h$) found by solving the fitting
problem for the three model Galactic potentials under
consideration and the masses of the Galactic components $m_b,
m_d,$ and $m_h$. The parameters $M_b, M_d,$ and $M_h$ are given in
units of the Galactic mass $M_{g}=2.325\times 10^7 M_\odot$;
$v^2_0/2$ in km$^2$ s${-2}$ is given instead of $M_h$ for model IV
(see the footnote to the table). The masses of the Galactic
components $m_b, m_d, m_h$ are given in units of the solar mass.
As has already been said above, the halo mass for models V and VI
when integrated over the entire Galaxy is $m_h=M_h$. The halo mass
within a sphere of radius 200 kpc determined by the object
velocity measurements is provided for model IV, because the halo
mass increases with increasing radius of the sphere $r$ as a
linear function of $r.$ The last row in the table gives the
residual between the model rotation curve and the measured
velocities (data) $\delta=\sqrt{\biggl(\sum_{i=1}^N
(V_{circ}(R_i)-\widetilde{V}_{circ}(R_i))^2\biggr)/N}$. The
next-to-last row provides the entropy of the difference between
the data and the model rotation curve. It gives an idea of the
degree of uniformity of the residual noise is:
 $E=-\frac{1}{N}\sum_{i=1}^N |\Delta_i|\ln(|\Delta_i|)$, where the
residual noise is
 $\Delta_i=V_{circ}(R_i)-\widetilde{V}_{circ}(R_i)$. The higher the
entropy, the more uniform the noise and, consequently, the better
the parameter fitting. As we see, model VI provides the smallest
residual $\delta=13.23$ km s$^{-1}$ and the greatest entropy of
the residual noise $E=-24.96$, i.e., the greatest degree of its
uniformity. This model is close to the Navarro--Frenk--White
(1997) model III, which we deem best from the viewpoint of the
accuracy of fitting to the data. In this case, $E=-24.51$ and
$\delta=13.1$ km s$^{-1}$ (see Table~1 in Bajkova and Bobylev
(2016)).

Table 2 lists the physical quantities calculated from the derived
parameters of the model potentials. These include the local disk
density $(\rho_\odot)_d$, the local dark matter density
$(\rho_\odot)_h,$ the local density of the entire matter
$\rho_\odot$, $K_{z=1.1}/2р G,$ the local surface density
$\Sigma_{1.1}$ and $\Sigma_{out}$, the two escape velocities from
the Galaxy $V_{esc}$ for $R=R_\odot$ and $R=200$~kpc, the linear
circular rotation velocity of the Sun $V_\odot,$, the Oort
constants $A$ and $B,$ and the Galactic mass $M_G$ for four radii
of the enclosing sphere.

The errors of all parameters and physical quantities in Tables~1
and 2, respectively, were determined through Monte Carlo
simulations using each time 100 independent realizations of random
velocity measurement errors by assuming the errors to obey a
normal distribution with zero mean and a fixed rms deviation.

The model Galactic rotation curves constructed for potentials IV,
V, and VI are presented in Fig.~1. The bulge, disk, and halo
contributions are given for each model.

In model IV, in accordance with (21), the circular velocity of the
halo increases monotonically with Galactocentric distance (see
Fig.~1). In this model it is apparently desirable to artificially
correct the halo density function at great distances
($R>200$~kpc), as is done in model I (Bajkova and Bobylev 2016).

Model V has the largest disk mass ($M_d$) compared to our other
models, as can be seen from Table~1. It follows from the last rows
in Table~2 that based on this model, we obtain the smallest
Galactic mass ($M_G$) among the other models. Models V and VI are
attractive in that both the circular velocity of the halo and the
overall rotation curve at distances greater than 100 kpc fall off
gently. Therefore, there is no need to artificially correct the
halo density function.

Note that we took the values justified by Irrgang et al. (2013) as
the target parameters in fitting $\rho_\odot$ and $K_{z=1.1}/2\pi
G$. Fitting these parameters basically led to fitting the function
$\rho_\odot(z)$ and, consequently, the surface density
$\Sigma_{1.1}$. The results of fitting to the function
$\rho_\odot(z)$ obtained by Irrgang et al. (2013) and considered
in this paper as a template are shown in Fig.~2. As can be seen
from the figure, all of the derived profiles are quite close to
one another, but model VI reproduces the required function
$\rho_\odot(z)$ with the highest accuracy.

Such local parameters of the rotation curve as the velocity
$V_\odot$ and the Oort constants $A$ and $B$ are well reproduced
by the models considered. In model V, however, $B$ is comparable
in absolute value to $A.$ Therefore, the rotation velocity in a
small segment near the Sun is nearly flat ($V_{circ}=const$), as
can be clearly seen from Fig.~1.

Interestingly, the escape velocity $V_{esc}$ ($R=200$~kpc) is
usually approximately half that at $R=R_\odot.$ However, for model
IV the parabolic velocity at $R=200$~kpc exceeds its value
calculated for $R=R_\odot.$

The Galactic mass within a sphere of radius 50 kpc,
$M_{50}\approx0.4\times10^{12}M_\odot,$ that we found based on
various model potentials is in good agreement with the results of
other authors. For example, Deason et al. (2012a) estimated
$M_{50}=(0.42\pm0.04)\times10^{12}M_\odot,$ from horizontal branch
giants; Williams and Evans (2015) found
$M_{50}=(0.45\pm0.15)\times10^{12}M_\odot$ from horizontal branch
giants from the SDSS (Sloan Digital Sky Survey) catalogue.

It is important to note that the Galactic mass estimates that we
obtained in this paper based on model VI (the best one in this
paper) are close to those based on model III (the best one in
Bajkova and Bobylev (2016)) and, consequently, are in good
agreement with the results of other authors that are displayed in
Fig. 4 from Bajkova and Bobylev (2016).

For an even subtler comparison of models III and VI we integrated
the orbits of two globular clusters (Lynga 7 and NGC 5053) from
the catalog by Kharchenko et al. (2013) based on the equations of
motion given in the Appendix to Irrgang et al. (2013). Figure 3
shows the Galactic orbits of the globular clusters Lynga 7 and NGC
5053 obtained by integration over 500 Myr and 5 Gyr into the past,
respectively. Table 3 gives the main characteristics of boxy
orbits: the average values of the minimum, $\langle
a_{min}\rangle$, and maximum, $\langle a_{max}\rangle$, distances
and the eccentricity $\langle e\rangle$. We can see good agreement
between the trajectories both for the cluster close to the
Galactic center and the short integration time interval (Lynga 7)
and for the fairly distant cluster and the long integration time
(NGC 5053).

It is interesting to compare our results with the result of other
authors. Although the present-day data suggest that the solar
velocity is $V_\odot=240$ km s$^{-1}$ and the solar Galactocentric
distance is $R_\odot=8.3$~kpc, many of the model potentials were
constructed for other values of $V_\odot$ and $R_\odot.$ As an
example, let us compare our results on the mass decomposition with
the results from Dauphole and Colin (1995). In this paper the
model potential V was used, and its parameters at
$R_\odot=8.0$~kpc and velocity $V_\odot=225$ km s$^{-1}$ were
found from the velocities of globular clusters in the range of
Galactocentric distances from 0 to 16 kpc. The following results
were obtained: $m_b=13.9\times10^9 M_\odot,
m_d=7.9\times10^{10}M_\odot, m_h=7.91\times10^{11}M_\odot$. The
local mass density was $\rho_\odot=0.143 M_\odot$ pc$^{-3}$; the
vertical force was $K_{z=1.1}/2\pi G=84~M_\odot$ pc$^{-2}$, which
differs tangibly from the present-day estimates that we used to
fit the parameters. By comparing the masses of the components with
those presented in Table~1, we can see that the bulge and disk
mass estimates are close to our estimates, while the halo masses
differ approximately by a factor of 2. This difference is
explained by the fact that the parameters were fitted using
different data. Our data on the velocities cover a wide range of
distances up to 200 kpc, while those from Dauphole and Colin
(1995) cover a range only up to 16 kpc, with our data showing a
considerable decrease in the circular velocities with increasing
distance $R$ at great distances. If we extend the model rotation
curve from Dauphole and Colin (1995) to large $R,$ then we will
not get the agreement with our data that is provided by our model
rotation curve. All of this suggests how important it is to use
the data in a wide range of Galactocentric distances to refine the
model Galactic potential.

 \section*{CONCLUSIONS}
We considered three three-component (bulge, disk, halo) model
Galactic potentials differing by the expression for the halo
potential. The observational data spanning the range of
Galactocentric distances $R$ from 0 to $\sim$200 kpc were used to
refine the parameters of these models. We relied on the
line-of-sight velocities of hydrogen clouds at the tangent points
and the data on 130 masers with measured trigonometric parallaxes
up to distances of about 20 kpc, and the average rotation
velocities from the review by Bhattacharjee et al. (2014) served
us as the data at greater distances.

In all models the bulge and disk potentials are represented by the
expressions from Miyamoto and Nagai (1975). The halo component is
described by a logarithmic Binney (1981) potential in model IV, is
represented by a Plummer (1911) sphere in model V, and is
described by a Hernquist (1990) potential in model VI. We solved
the fitting problem by taking into account the additional
constraints imposed on (a) the local matter density $\rho_\odot$
and (b) the force $K_{z=1.1}$ acting perpendicularly to the
Galactic plane. As a result, we obtained the model potentials that
described a stellar system consistent with the physical
characteristics of visible matter in the Galaxy known from
observations.

Model VI looks best among the three models considered in this
paper from the viewpoint of the accuracy of fitting the model
rotation curves to the measurements and to the local matter
density and the force acting perpendicularly to the Galactic
plane. At the same time, the Navarro–Frenk–White model III is the
best one among the six models we considered, the first three of
which were considered in our previous paper (Bajkova and Bobylev
2016). It should be noted that models III and VI are close, which
we showed, in particular, using the integration of the orbits of
two globular clusters as an example.

The Galactic masses within a sphere of radius 50 kpc are close in
models IV--VI or, more specifically,
 $M_{50}=(0.409\pm0.020)\times10^{12}M_\odot$ in model IV and
 $M_{50}=(0.417\pm0.034)\times10^{12}M_\odot$ in models V and VI. The differences
between the models increase with increasing radius of the sphere.
For example, the Galactic mass within a sphere of radius 200 kpc
is
 $M_{200}=(1.395\pm0.082)\times10^{12}M_\odot$ in model IV,
 $M_{200}=(0.469\pm0.038)\times10^{12}M_\odot$ in model V, and
 $M_{200}=(0.641\pm0.049)\times10^{12}M_\odot$ in model VI.

 \medskip
This work was supported by the ``Transient and Explosive Processes in
Astrophysics'' Program P--7 of the Presidium of the Russian
Academy of Sciences.

 \bigskip\medskip{REFERENCES}\medskip{\small

 1. C. Allen and A. Santill\'an, Rev.Mex. Astron. Astrofis. 22, 255 (1991).

 2. C.A.L. Bailer-Jones, Astron. Astrophys. 575, 35 (2015).

 3. A.T. Bajkova and V.V. Bobylev, Astron. Lett. 42, 567 (2016).

 4. P. Bhattacharjee, S. Chaudhury, and S. Kundu, Astrophys. J. 785, 63 (2014).

 5. J. Binney, Mon. Not. R. Astron. Soc. 196, 455 (1981).

 6. V.V. Bobylev and A.T. Bajkova, Astron. Lett. 39, 809 (2013).

 7. R.A. Burns, H. Imai, T. Handa, T. Omodaka, A. Nakagawa, T. Nagayama,
and Y. Ueno, Mon. Not. R. Astron. Soc. 453, 3163 (2015).

8. W.B. Burton and M.A. Gordon, Astron. Astrophys. 63, 7 (1978).

9. M.A. Butenko and A.V. Khoperskov, Vestn. Volgogr. Univ., Ser.
1: Mat. Fiz., No. 1 (20), 61 (2014).

10. R. Capuzzo-Dolcetta and G. Fragione, Mon. Not. R. Astron. Soc.
454, 2677 (2015).

11. Y.K. Choi, K. Hachisuka, M.J. Reid, Y. Xu, A. Brunthaler, K.M.
Menten, and T.M. Dame, Astrophys. J. 790, 99 (2014).

12. B. Dauphole and J. Colin, Astron. Astrophys. 300, 117 (1995).

13. A.J. Deason, V. Belokurov, N.W. Evans, and J. An, Mon. Not. R.
Astron. Soc. 424, L44 (2012a).

 14. A.J. Deason, V. Belokurov, N.W. Evans, S.E. Koposov, R.J.
Cooke, J. Penarrubia, C.F.P. Laporte, M. Fellhauer, et al., Mon.
Not. R. Astron. Soc. 425, 2840 (2012b).

15. G.N. Dremova, V.V. Dremov, V.V. Orlov, A.V. Tutukov, and K.S.
Shirokova, Astron Rep. 59, 1019 (2015).

 16. G.M. Eadie, W.E. Harris, and L.M. Widrow, Astrophys. J. 806,
54 (2015).

17. N.W. Evans, Mon. Not. R. Astron. Soc. 260, 191 (1993).

18. M. Fellhauer, V. Belokurov, N.W. Evans, M.I. Wilkinson, D.B.
Zucker, G. Gilmore, M.J. Irwin, D.M. Bramich, et al., Astrophys.
J. 651, 167 (2006).

19. O.Y. Gnedin, W.R. Brown, M.J. Geller, and S.J. Kenyon,
Astrophys. J. 720, L108 (2010).

20. G.A. Gontcharov, A.T. Bajkova, P.N. Fedorov, and V.S.
Akhmetov, Mon. Not. R. Astron. Soc. 413, 1581 (2011).

21. K. Hachisuka, Y.K. Choi, M.J. Reid, A. Brunthaler, K.M.
Menten, A. Sanna, and T.M. Dame, Astrophys. J. 800, 2 (2015).

22. L. Hernquist, Astrophys. J. 356, 359 (1990).

23. A. Irrgang, B. Wilcox, E. Tucker, and L. Schiefelbein, Astron.
Astrophys. 549, 137 (2013).

24. A. Just, P. Berczik,M. I. Petrov, and A. Ernst, Mon. Not. R.
Astron. Soc. 392, 969 (2009).

25. J.S. Kalirai, J. Anderson, H.B. Richer, I.R. King, J.P.
Brewer, G. Carraro, S.D. Davis, G.G. Fahlman, B.M.S. Hansen, et
al., Astrophys. J. 657, L93 (2007).

26. I.D. Karachentsev, O.G. Kashibadze, D.I. Makarov, and R.B.
Tully, Mon.Not. R. Astron. Soc. 393, 1265 (2009).

27. N.V. Kharchenko, A.E. Piskunov, E. Schilbach, S. Roeser, and
R.-D. Scholz, Astron. Astrophys. 558, A53 (2013).

28. G.G. Kuzmin, Astron. Zh. 33, 27 (1956).

29. M. Miyamoto and R. Nagai, Publ. Astron. Soc. Jpn. 27, 533
(1975).

30. K. Motogi, K. Sorai, M. Honma, T. Hirota, K. Hachisuka, K.
Niinuma, K. Sugiyama, Y. Yonekura, and K. Fujisawa, Publ. Astron.
Soc. Jpn. 68, 69 (2016).

31. J.F. Navarro, C.S. Frenk, and S.D.M. White, Astrophys. J. 490,
493 (1997).

32. H.C. Plummer, Mon. Not. R. Astron. Soc. 71, 460 (1911).

33. M.J. Reid, K.M. Menten, A. Brunthaler, X.W. Zheng, T.M. Dame,
Y. Xu, Y. Wu, B. Zhang, et al., Astrophys. J. 783, 130 (2014).

 34. A. R{u}\u{z}i\u{c}ka, C. Theis, and J. Palou\u{s}, Astrophys.
J. 725, 369 (2010).

35. M. Sato, Y.W. Wu, K. Immer, B. Zhang, A. Sanna, M.J. Reid,
T.M. Dame, A. Brunthaler, and K.M. Menten, Astrophys. J. 793, 72
(2014).

36. D. Veras and N.W. Evans, Mon. Not. R. Astron. Soc. 430, 403
(2013).

37. M.I. Wilkinson and N.W. Evans, Mon. Not. R. Astron. Soc. 310,
645 (1999).

38. A.A. Williams and N.W. Evans, Mon. Not. R. Astron. Soc. 454,
698 (2015).

39. Y.W. Wu, M. Sato, M.J. Reid, L. Moscadelli, B. Zhang, Y. Xu,
A. Brunthaler, K.M. Menten, T.M. Dame, and X.W. Zheng, Astron.
Astrophys. 566, 17 (2014).

40. A.V. Zasov and K.A. Postnov, General Astrophysics (Vek 2,
Fryazino, 2006) [in Russian].

 }

 \end{document}